\documentclass[aps,prx,reprint,superscriptaddress]{revtex4-1}

\usepackage{amsmath}
\usepackage{xcolor}
\usepackage{bbold}
\usepackage{graphicx}
\usepackage{mathtools}
\usepackage{dsfont}
\usepackage{accents}

\begin{document}


\title{K-spin Hamiltonian for quantum-resolvable Markov decision processes}


\author{Eric B. Jones}
\email[]{ebjones@mymail.mines.edu}
\email[]{Eric.Jones@nrel.gov}
\affiliation{National Renewable Energy Laboratory, Golden, CO 80401, USA}
\affiliation{Department of Physics, Colorado School of Mines, Golden, CO 80401, USA}

\author{Peter Graf}
\affiliation{National Renewable Energy Laboratory, Golden, CO 80401, USA}

\author{Eliot Kapit}
\affiliation{Department of Physics, Colorado School of Mines, Golden, CO 80401, USA}

\author{Wesley Jones}
\affiliation{National Renewable Energy Laboratory, Golden, CO 80401, USA}


\date{\today}

\begin{abstract}
The Markov decision process is the mathematical formalization underlying the modern field of reinforcement learning when transition and reward functions are unknown. We derive a pseudo-Boolean cost function that is equivalent to a K-spin Hamiltonian representation of the discrete, finite, discounted Markov decision process with infinite horizon. This K-spin Hamiltonian furnishes a starting point from which to solve for an optimal policy using heuristic quantum algorithms such as adiabatic quantum annealing and the quantum approximate optimization algorithm on near-term quantum hardware. In proving that the variational minimization of our Hamiltonian is equivalent to the Bellman optimality condition we establish an interesting analogy with classical field theory. Along with proof-of-concept calculations to corroborate our formulation by simulated and quantum annealing against classical Q-Learning, we analyze the scaling of physical resources required to solve our Hamiltonian on quantum hardware.
\end{abstract}


\maketitle

\section{Introduction \label{intro}}
Alongside supervised and unsupervised learning, reinforcement learning (RL) constitutes one of the main pillars of modern attempts to create artificial intelligence \cite{sutton2018reinforcement}. RL describes a set of techniques, anchored in the mathematical formalism of the Markov decision process (MDP), whereby an autonomous agent learns how to behave according to an optimal policy within an environment in order to maximize some notion of expected future rewards derived from said environment by repeatedly interacting with it. When the agent finds itself in some state of the environment, it chooses an action that causes it to transition to a subsequent state of the environment and simultaneously be given a reward (or penalty). The agent's behavioral policy is therefore typically regarded as a probability distribution over state-action pairs, which tells the agent which action(s) is (are) best to take in each state. RL has begun to see use in a wide variety of applications including recommendation systems and optimal control, and perhaps the most stark demonstration of the computational utility of RL has been its role in enabling a computer to beat expert human players at the game of Go \cite{golovin2004reinforcement,sutton1992reinforcement,silver2018general}.

While RL is a novel and powerful algorithmic paradigm, its implementation currently remains largely tied to computational resources that are dictated by classical physics at the hardware level. Given that the space of possible policies for an agent to enumerate is often combinatorially large, such as in the instance of Go, there is good reason to suspect that the particular type of parallelism afforded by quantum hardware might help an agent in finding an optimal policy more quickly. In fact, an existing body of literature already explores quantum formulations of RL in situations where either the agent or environment, or both, are regarded as quantum mechanical, and in some instances a quantum speedup has been established (see for example \cite{dong2008quantum,dunjko2017advances,lamata2017basic,briegel2012projective,paparo2014quantum,dunjko2016quantum,dunjko2015quantum,neukart2018quantum}). To our knowledge however, there does not yet exist a formulation of a quantum MDP with input from a classical environment that is amenable to solution directly on near-term, noisy intermediate-scale quantum hardware via existing quantum optimization algorithms.

In this work, we therefore establish what we regard to be the simplest and most straightforward connection between the mathematical formalism of RL and that of quantum computing. Namely, we consider a ``quantum-accelerated'' agent interacting with a classical environment. Moreover, we suppose that the agent possesses a model of its environment such that the quantum MDP consists only of the agent solving for an optimal policy using quantum hardware. Strictly speaking, this formulation in and of itself does not constitute ``quantum RL'' because the agent already has a model of its environment. However, paired with a classical routine where the agent learns the transition and reward rules for its environment, our quantum-resolved MDP may be thought of as a (potentially hybrid) quantum subroutine embedded in a larger hybrid quantum-classical RL algorithm. 

Note that all three variants of the general MDP (finite horizon, infinite horizon discounted, and infinite horizon averaged cost) have been shown to be P-complete \cite{papadimitriou1987complexity}. This means roughly that while MDPs are in principle efficiently solvable in polynomial time, they are not categorically efficiently parallelizable \cite{greenlaw1995limits}. Moreover, a general MDP suffers from the ``curse of dimensionality'', meaning that its attendant memory requirements grow exponentially in the dimension of the state space. In practice, a conventional computer can solve MDPs involving millions of states \cite{sutton2018reinforcement}.

Our approach is motivated by the observation that a majority of near-term quantum algorithms such as adiabatic quantum annealing and the quantum approximate optimization algorithm are concerned with the solution of discrete, combinatorial optimization problems recast as K-spin Hamiltonians of the form \cite{farhi2014quantum,denchev2016computational}
\begin{equation} \label{eq:gen_ham}
\hat{H} = - \sum_{k=1}^{K} \sum_{j_1 \ldots j_k = 1}^N J_{j_1 \ldots j_k} \hat{Z}_{j_1} \ldots \hat{Z}_{j_k}.
\end{equation}
As such, we identify the policy with discrete variables to be mapped to spins (and thus qubits) and a sum over Q-functions with a K-spin Hamiltonian. Since it is well-known that finding the ground state of a K-spin Hamiltonian is in the complexity class NP-complete through its polynomial-time order reduction to the Ising spin Hamiltonian, it may at first glance look odd to map a problem from P-complete into a formulation, which ostensibly resides in NP-complete \cite{barahona1982computational,rosenberg1975reduction}. Our primary justification is that certain MDP generalizations, such as the partially observable MDP (POMDP), are harder than the MDP- for example POMDP is PSPACE-complete \cite{papadimitriou1987complexity}. The present work provides a platform from which to explore K-spin Hamiltonians as an approximation scheme for problems such as POMDP. In addition, while \textit{arbitrary} instances of the K-spin problem are NP-complete there can be \textit{specific} instances of the K-spin problem, which are soluble in polynomial time \cite{zintchenko2015local,cipra2000ising}. It is possible that the K-spin Hamiltonian derived in the present work inherits its polynomial-time solubility from the dynamic programming formulation from which it is derived. Suggesting otherwise however, is the fact that when reformulated into Ising spin form, the $k$-regular $k$-XORSAT problem has been shown to take exponential time to solve by adiabatic quantum annealing when $k=3$ while Gaussian elimination allows polynomial-time solubility classically \cite{farhi2012performance,patil2019obstacles,bapst2013quantum}. We therefore do not consider our quantum MDP a panacea for all large-order polynomial-time MDPs. Rather, it furnishes an interesting connection between two disparate fields in computational science while allowing for quantum parallelization of a problem, which is not efficiently parallelizable on a classical computer (unless $P = NC$).

\section{Markov Decision Process \label{MDP}}
The basic mathematical formalism that describes an autonomous agent interacting with its environment is the MDP. Our treatment follows from the standard text by Sutton and Barto \cite{sutton2018reinforcement}. We take the discrete, finite, discounted MDP with infinite horizon to be a 5-tuple $(S,A,P,R,\gamma)$ where $S$ is a finite set of discrete states in which the agent may find itself, $A$ is a finite set of discrete actions that the agent may carry out, $P$ is a set of transition probabilities of the form $P_{ss'}^a$, which describe the probability that an agent will transition to state $s'$ from state $s$ given that it has taken action $a$, $R$ is a set of rewards of the form $R_{ss'}^a$, which describe the reward that an agent receives upon moving from state $s$ to state $s'$ under the action $a$, and $\gamma$ is the discount factor, which gauges the degree to which rewards farther into the future are prioritized less than more immediate ones.

The goal of an agent moving around in an environment is to learn an optimal policy $\pi_{sa}$ such that when it finds itself in the state $s$, $\pi_{sa}$ is the probability that the agent should take the action $a$ in order to maximize its expected future discounted reward. Formally, there are two quantities that represent an expected future discounted reward under the policy $\pi$: the value function
\begin{align}\label{eq:val_fun}
V_s[\pi] &= \mathbb{E}_{\pi}[G_t | S_t = s],
\end{align}
which describes the future discounted reward expected by virtue of being in state $s$ at time $t$, and the action-value function (also known as the Q-function)
\begin{align}\label{eq:act_val_fun}
Q_{sa}[\pi] &= \mathbb{E}_{\pi}[G_t | S_t = s, A_t=a] ,
\end{align}
which describes the future discounted reward expected by choosing action $a$ when starting in state $s$ at time $t$. In either case, the quantity
\begin{equation}
G_t = \sum_{k=0}^{\infty} \gamma^k R_{t+k+1}
\end{equation}
is the discounted return expected by the agent from time $t$ onward.
\begin{figure}
\includegraphics[width=1\linewidth]{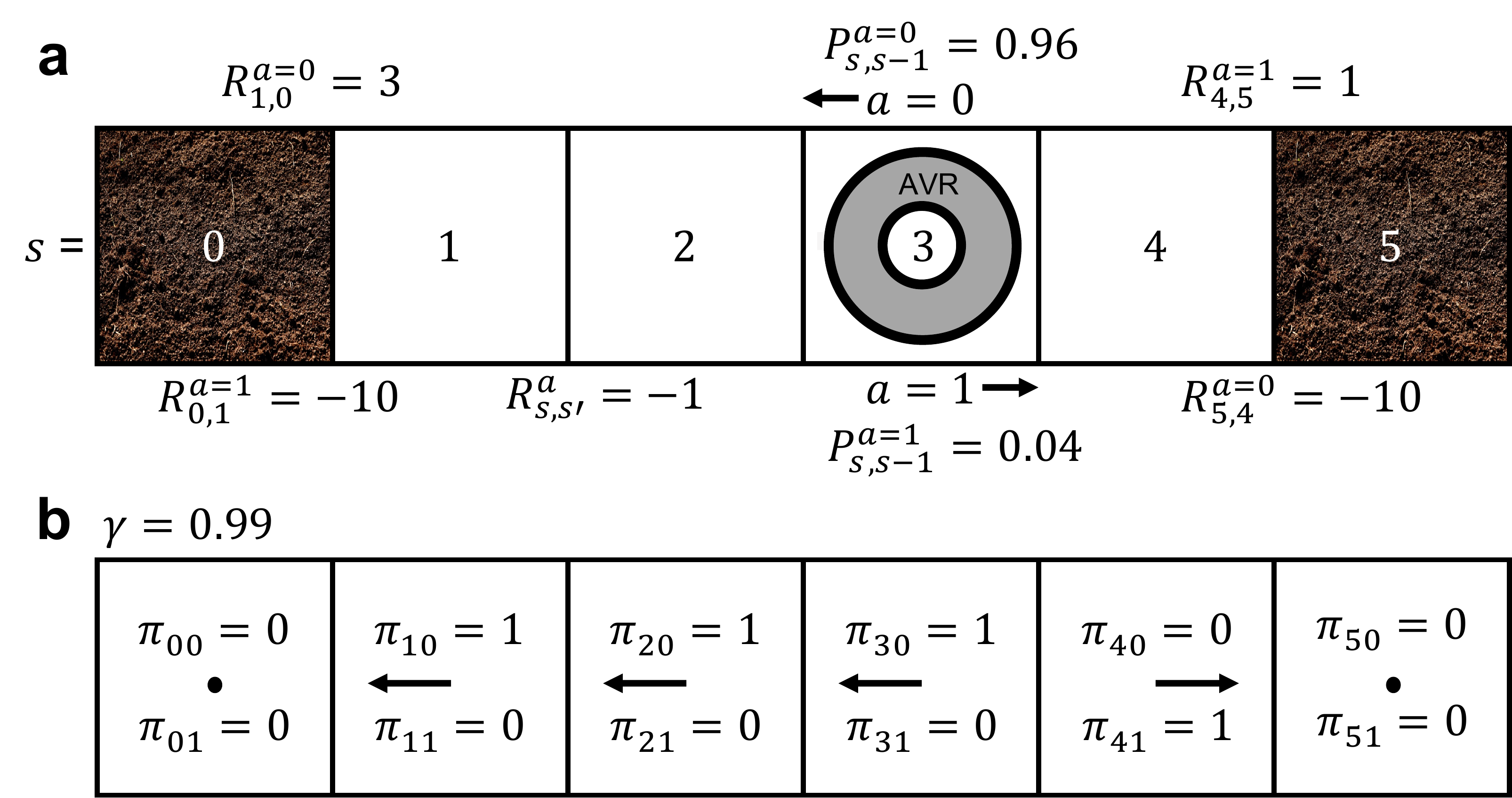}
\caption{The Markov decision process. \textbf{a} AVR in a hallway trying to find the larger dirt pile to clean formulated as an MDP. \textbf{b} Optimal policy at discount factor $\gamma = 0.99$. \label{roomba}}
\end{figure}
Intuitively, the agent wants to collect as many of the largest, positive-valued $R_{ss'}^a$ as possible and as few negative-valued ones over some effective time horizon dictated by the value of $\gamma$. If the sets $P$ and $R$ are known a priori the agent is said to have a model of its environment. If not, the agent must learn features of its environment implicitly or through a training simulation, which are situations encompassed by the MDP. To make these concepts concrete consider the particular example of an autonomous vacuuming robot (AVR) in a hallway trying to find the larger of two dirt piles to clean as displayed in Fig.~\ref{roomba}a. The states accessible to the AVR, $s\in S = \{0, \ldots ,5\}$, are the six floor tiles, and the action space is $A=\{0,1\}$ with $a=0$ corresponding to moving left and $a=1$, right. The terminal states, $s=0$ and $s=5$, correspond to dirt piles that the AVR wants to find and clean. As such, if it encounters a dirt pile, the AVR receives the rewards $R_{4,5}^1=1$ (the smaller pile) and $R_{1,0}^0 = 3$ (the larger pile). If we suppose that each AVR is responsible for finding and cleaning only one dirt pile, then we can entice the AVR to stay at the dirt pile it has found by introducing a steep penalty for moving away $R_{5,4}^0 = R_{5,5}^1 = R_{0,1}^1 = R_{0,0}^0 = -10$. Note that one could also immobilize the AVR with $P_{0,0}^a = P_{5,5}^a \sim 1$. Each step in the interior of the hallway, say between $s=2$ and $3$, incurs a penalty $R_{s,s' \in interior}^a =-1$ so that the AVR does not dawdle. We specify that this process is slightly stochastic in that $P_{s,s-1}^0 = P_{s,s+1}^1 = 0.96$ and $P_{s,s-1}^1 = P_{s,s+1}^0 = 0.04 \: \: \forall s$. In this context, stochasticity could reflect a tendency of the AVR to occasionally misfire and move in the wrong direction. We make this designation because the deterministic MDP is even easier (NC complexity class) than the general, stochastic MDP \cite{papadimitriou1987complexity}. At the edges of the hallway the AVR sees relatively hard boundary conditions via $P_{0,0}^0 = P_{5,5}^1 = 0.96$. We also regard the policy, $\pi$, as a binary indicator variable such that $\pi_{sa}=1$ if the agent takes action $a$ in state $s$ and $\pi_{sa}=0$ if it does not. Such a policy is called a deterministic policy. However, the K-spin Hamiltonian derived in Sec.~\ref{Kspin} is valid for non-deterministic policies as well since any probabilistic policy expressed as a fixed-precision or floating-point number has a linear expansion in binary variables and thus can be mapped linearly to qubit degrees of freedom \cite{yates2009fixed}. The use of deterministic policy variables simply allows a more economical use of qubit resources. We remark that this environment is equivalent to a 1-dimensional version of the classic ``Gridworld'' environment encountered in the RL literature \cite{sutton2018reinforcement}. 

As established, one way for an agent to understand the quality of its current policy is by calculating the action-value function $Q_{sa}[\pi]$ in Eq.~\ref{eq:act_val_fun}. For a given policy, the action-value function returns the expected future discounted reward for taking action $a$ in state $s$ and then subsequently following $\pi$. An optimal policy is one for which $Q_{sa}[\pi]$ is maximized $\forall \: \: s, a$. One of the central results of the dynamic programming treatment of MDPs is that the action-value function obeys a Bellman equation
\begin{equation}\label{Q_Bell}
Q_{sa}[\pi] = \sum_{s'} P_{ss'}^a \Bigg(R_{ss'}^a + \gamma \sum_{a'} \pi_{s'a'} Q_{s'a'}[\pi] \Bigg).
\end{equation}
We use Eq. \ref{Q_Bell} as the starting point from which to derive a K-spin Hamiltonian for a general discrete, finite, discounted MDP over an infinite horizon.

\section{MDP K-spin Hamiltonian \label{Kspin}}
\begin{table}[] 
\caption{ Analogy between classical field theory and the MDP. \label{analogy}}
\begin{ruledtabular}
\begin{tabular}{ | c || c | c | }
Object & Classical Field Theory & MDP  \\
\hline\hline
Coordinate & $x=(x^0,x^1,x^2,x^3)$ & $\mu = (s,a)$ \\
\hline
Field & $\Phi(x)$ & $\pi_{\mu}$ \\
\hline
Density & $\mathcal{L}(\Phi(x),x)$ & $Q_\mu[\pi]$ \\
\hline
Functional & $S[\Phi] = \int d^4x \mathcal{L}(\Phi(x),x)$ & $H[\pi] = \sum_{\mu} Q_\mu [\pi]$ \\
\hline
Stationarity & $\delta S = 0$ & $\delta H =0$ \\
\hline
Euler-Lagrange & $\delta \mathcal{L} / \delta \Phi = 0$ & $\delta Q / \delta \pi = 0$
\end{tabular}
\end{ruledtabular}
\end{table}

One can express the action-value function purely as a functional of the policy by repeatedly substituting the left-hand side of Eq. \ref{Q_Bell} into the right-hand side of Eq. \ref{Q_Bell}. The result is
\begin{align}
&Q_{sa}[\pi] = \sum_{s'} P_{ss'}^a R_{ss'}^a + \sum_{s',a'} \pi_{s'a'} \Bigg( \gamma \sum_{s''} P_{ss'}^a P_{s's''}^{a'} R_{s's''}^{a'} \Bigg) \nonumber \\
&+ \sum_{s',a',s'',a''} \pi_{s'a'} \pi_{s''a''} \Bigg( \gamma^2 \sum_{s'''} P_{ss'}^a P_{s's''}^{a'} P_{s''s'''}^{a''} R_{s''s'''}^{a''} \Bigg) \nonumber \\
&+ \ldots.
\end{align}
$Q$ is now a functional of the policy. A stationary point of a functional occurs whenever its variation with respect to its argument vanishes \cite{dreyfus1965dynamic}. Therefore finding the optimal policy that maximizes the Q-function everywhere is equivalent to the variational condition
\begin{equation}\label{Q_var}
\frac{\delta Q_{sa} [\pi]}{\delta \pi_{\bar{s}{\bar{a}}}} = 0
\end{equation}
for an arbitrary variation of $Q$ with respect to the policy. From here it is straightforward to define a Hamiltonian functional as
\begin{equation}\label{Ham_func}
H[\pi] \equiv - \sum_{s,a} Q_{s,a}[\pi].
\end{equation}
Recall here that $\pi_{sa} \in \{0,1\}$ and therefore has a simple relationship to single-qubit spin operators via $\pi_{sa} \rightarrow (\hat{\mathds{1}}_{sa}-\hat{Z}_{sa})/2$. So, Eq.~\ref{Ham_func} is equivalent to a K-spin Hamiltonian where $\gamma^K$ multiplies the largest order in the infinite expansion that needs to be taken into account in order to find the optimal policy.

The Hamiltonian in Eq. \ref{Ham_func} is amenable to simulation on a quantum computer either via adiabatic or variational preparation. What one needs to check is whether minimizing $H$ with respect to $\pi$ is equivalent to maximizing $Q_{sa}$ for each $s$ and $a$. For this purpose, consider an arbitrary variation of the Hamiltonian with respect to the policy. Minimization of the Hamiltonian corresponds to
\begin{align}
\frac{\delta H [\pi]}{\delta \pi_{\bar{s}{\bar{a}}}} &= 0 \nonumber \\
&= - \sum_{s,a} \frac{\delta Q_{sa} [\pi]}{\delta \pi_{\bar{s}{\bar{a}}}}.
\end{align}
But since the variation is arbitrary, each term in the sum needs to vanish identically leading to Eq. \ref{Q_var} for all $s, a, \bar{s}, \bar{a}$. 

There is an interesting analogy to classical field theory in the present discussion shown in Table~\ref{analogy}. Each state-action pair $\mu=(s,a)$ plays the role of a spacetime coordinate over which a field $\pi_{\mu}$ is defined. The action-value function can loosely be thought of similar to a Lagrangian density. The Hamiltonian functional plays a role analogous to the action functional of field theory and its variation leads to an ``Euler-Lagrange'' equation for $Q$. We note that there exist path integral approaches to both optimal control and reinforcement learning, and Markov chains have been explored as tools in field theory \cite{kappen2005path,theodorou2010generalized,dynkin1983markov}. It would be an interesting direction for future work to explore the possible connections between our K-spin theory and these previous approaches. In addition, we emphasize that by expressing the MDP in the form of Eq.~\ref{Ham_func} (along with the conditional probability normalization below in the second line of Eq.~\ref{eq:bool_Ham}), a quantum optimization heuristic will attempt to produce the optimal policy for the entire state-action space at once by virtue of the fact that the Hamiltonian contains in its parameters a model for the entire environment. In the absence of a known model, such as in the instance where the state-action space is too large to allow tabulation of all transition and reward functions or where such functions simply are not known a priori, we expect that one should still be able to write down a Hamiltonian whose ground state infers a good policy by sampling transition and reward functions from across the state-action space. We leave a demonstration of this concept for future work as well, since it is beyond the scope of the present paper.

Explicitly then, the Hamiltonian we may use to solve an MPD for an optimal policy is
\begin{align}\label{eq:bool_Ham}
H[\pi] &= - \sum_{k=0}^{\infty} \sum_{\mu_1 \ldots \mu_k=1}^{|S\times A|} J_{\mu_1 \ldots \mu_k} \pi_{\mu_1} \ldots \pi_{\mu_k} \\
&+ M \sum_{s_1=1}^{|S|} \Bigg( \sum_{a_1=1}^{|A|} \pi_{s_1 a_1} - 1 \Bigg)^2 \nonumber
\end{align}
where we have made the identifications $s (a) \rightarrow s_0 (a_0)$, $s' (a') \rightarrow s_1 (a_1)$, etc. and
\begin{align}\label{eq:coup_coeffs}
J_{\mu_1 \ldots \mu_k} &\equiv J_{(s_1, a_1) \ldots (s_k, a_k)} \nonumber \\
&\equiv \gamma^k \sum_{s_0,a_0,s_{k+1}} P^{a_0}_{s_0 s_1} \ldots P^{a_k}_{s_k s_{k+1}} R^{a_k}_{s_k s_{k+1}}.
\end{align}
Eq.~\ref{eq:bool_Ham} includes a penalty function with strength $M$ in the second line due to the fact that the policy $\pi_{sa}$ is generally a probability distribution conditioned on the state variable $s$ and so must equal unity when summed over all actions ($\sum_{a_1} \pi_{s_1 a_1} =1 \: \: \forall \: \: s_1 $). Also, note that the $k=0$ term involves a slight abuse of notation and is actually just a constant offset to the Hamiltonian, which otherwise displays the same form as Eq.~\ref{eq:gen_ham} with $K \rightarrow \infty$. The quantities defined in Eq.~\ref{eq:coup_coeffs} are $k-$qubit coupling coefficients, which gauge how many terms in the Hamiltonian need to be retained in order to solve for an optimal policy. Even though $\gamma^k$ monotonically decreases with $k$ since $0 < \gamma < 1$,
it is not clear a priori that Eq.~\ref{eq:gen_ham} is convergent for a general MDP problem instance. However, even if Eq.~\ref{eq:gen_ham} is an asymptotic series in some instances, it may still provide useful solutions as is well known within the context of perturbative quantum field theory \cite{peskin2018introduction}. We take $k=K$ as the order at which one is practically able to truncate the series.

In terms of hardware requirements, a larger $K$ will require higher gate complexity and depth for gate-model machines and larger numbers of ancillary qubits for annealer-model machines in order to perform order-reduction \cite{pedersen2019native,boros2002pseudo}. Gate-model machines may also use ancillary variables in order to reduce gate complexity. In addition, if $K$ is chosen to be too small with respect to the decay of $\gamma^k$ then the ground state of the Hamiltonian will fail to exhibit some long-range correlations that might be essential to finding an optimal policy at every state-action pair. We will discuss hardware resource requirements for accurate policy determination in more detail in Sec.~\ref{Hardware}.
\begin{figure}[]
\includegraphics[width=\linewidth]{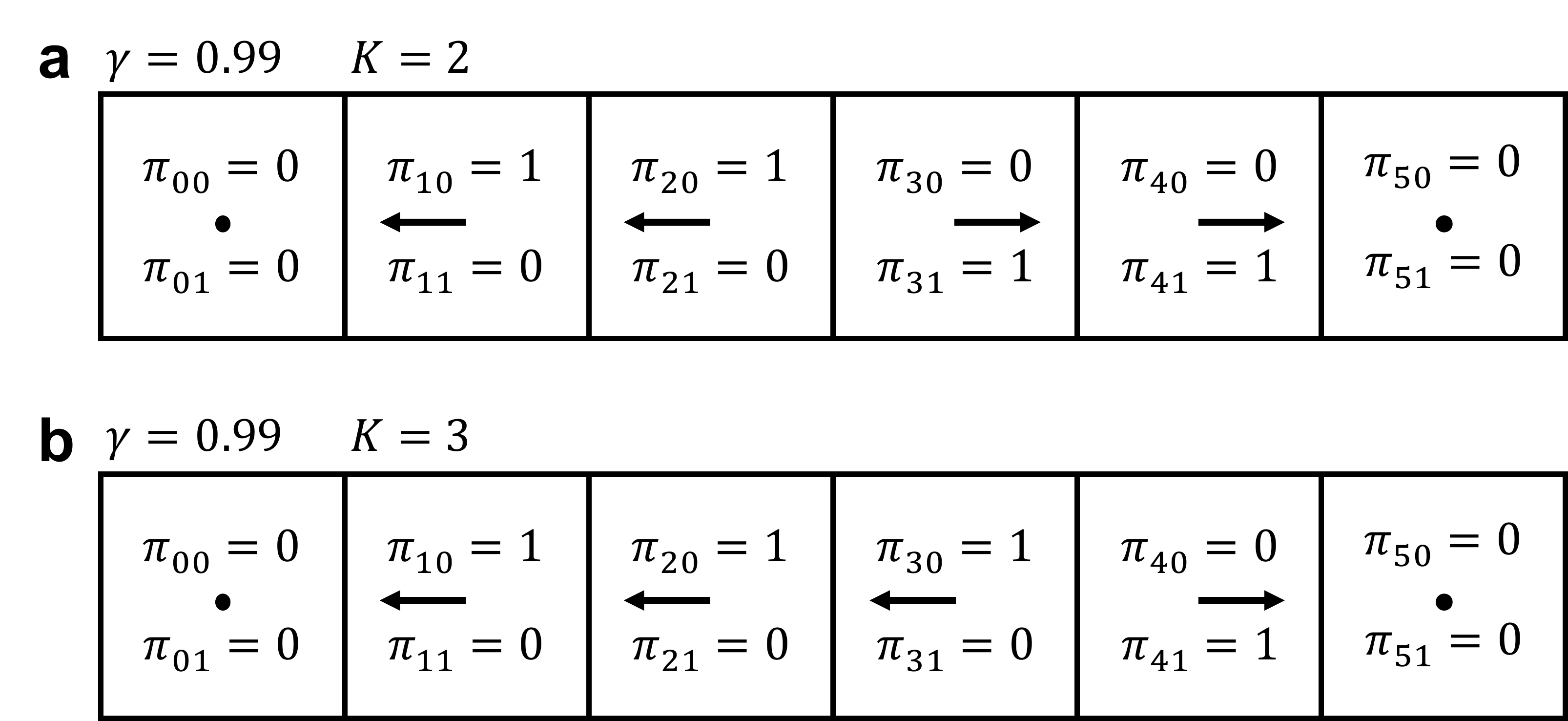}
\caption{Effects of prematurely truncating the K-spin Hamiltonian. \textbf{a} At $\gamma=0.99$ and $K=2$ for a 6-tile state space and asymmetric terminal rewards specified in Sec.~\ref{MDP}, the optimal policy is erroneously symmetric. Note however, that this policy is the true ground state for $\gamma=0.8$ and $K=2$. \textbf{b} When $K$ is raised to $3$, the AVR accurately identifies longer-term rewards and moves left at the fourth tile from left instead of moving right. \label{roomba_asym}}
\end{figure}
Fig.~\ref{roomba}b shows the optimal policy found by solving for the ground state of $H$ for the AVR-dirt scenario described in Sec.~\ref{MDP} on the D-Wave 2000Q processor and validated by simulated annealing and classical Q-Learning run in an OpenAI Gym environment \cite{brockman2016openai}. For the Q-Learning hyperparameters, we use $\alpha=0.1$ (learning rate), $\epsilon = 0.1$ (convergence threshold), and $N_{ep}=20,000$ (number of training episodes). A more detailed discussion of quantum and simulated annealing and their hyperparameters will follow in Sec.~\ref{Time}. The strength of the penalty function in Eq.~\ref{eq:bool_Ham} is set to $M=3$, the discount factor is set to $\gamma=0.99$, and the Hamiltonian is truncated at $K=3$ and then reduced to quadratic order, which corresponds to a quadratic unconstrained binary optimization (QUBO) or Ising model. The procedure for Hamiltonian order reduction will be discussed in Sec.~\ref{Hardware}, but we remark here that we set the penalty strength required for order reduction to $M_{OR}=5$. Fig.~\ref{roomba_asym} shows how premature truncation adversely affects solution quality for the same environment. In Fig.~\ref{roomba_asym}a, truncation at $K=2$ erroneously leads to a policy that is symmetric in the state space due to the fact that terms, which couple policy variables in the fourth tile from left to the left-most tile are neglected. By truncating instead at $K=3$, the AVR becomes aware of longer-term rewards and therefore identifies the correct policy.  Due to this observation, we turn to an analysis of the scaling of spatial resources (i.e. numbers of qubits and Hamiltonian parameters) in Sec.~\ref{Hardware} before assessing the scaling in time complexity of the K-spin MDP in Sec.~\ref{Time}.
\section{Hardware Requirements for the K-Spin MDP \label{Hardware}}

\begin{figure}
\includegraphics[width=\linewidth]{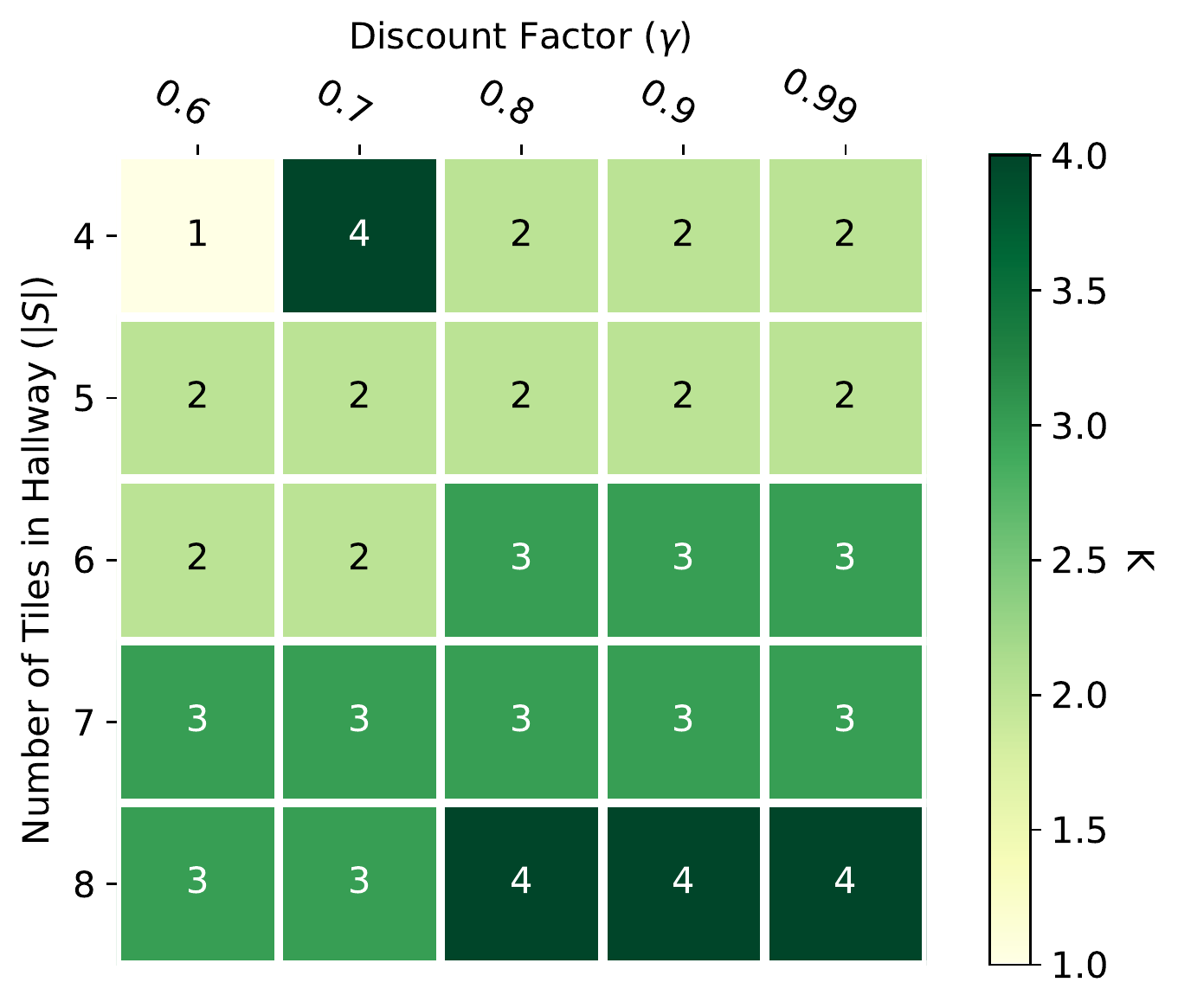}
\caption{Lowest order ($K$) at which Eq.~\ref{eq:bool_Ham} can be truncated and still find the unique, optimal policy as a function of system size ($|S|$) and discount factor ($\gamma$) for the environment described in Sec.~\ref{MDP}. The truncation order scales roughly as $K \sim |S|/2$. \label{K_heatmap}}
\end{figure}

Given that the order ($k=K$) at which Eq.~\ref{eq:bool_Ham} is truncated affects whether or not the ground-state solution equates to the optimal policy, a natural question to ask is: how many terms in the Hamiltonian need to be retained? This question is also important since either the number of ancilla qubits or gate complexity or both will grow as a function of $K$. Fig.~\ref{K_heatmap} shows the lowest truncation order ($K$) at which the optimal policy was found as the unique ground state of Eq.~\ref{eq:bool_Ham} as a function of system size ($|S|$) and discount factor ($\gamma$) by simulated annealing. For our hallway environment, we find that the truncation order scales roughly as $K\sim |S|/2$ for typical values of the discount factor such as $\gamma=0.9$, but for a general environment we expect that this scaling will depend upon the form of the transition and reward probabilities $P$ and $R$ as well as the dimensions of the state and action spaces $S$ and $A$. The aberrant $K=4$ result observed at $(|S|=4,\gamma=0.7)$ is a consequence of the small size of the state space combined with a small discrepancy in how the boundary conditions are implemented classically. Boundary conditions matter less as the state space becomes larger. Otherwise, Fig.~\ref{K_heatmap} is encouraging for two reasons. First, while Eq.~\ref{eq:bool_Ham} formally involves an infinite number of terms, Fig.~\ref{K_heatmap} confirms that in practice, only a finite number of terms need to be retained in order to find an optimal policy. This situation is similar to the dynamic programming treatment for MDPs wherein only a finite number of policy iterations are needed in order to find an optimal policy \cite{sutton2018reinforcement}. And second, we see that as the discount factor ($\gamma$) is varied, our Hamiltonian's ground state is able to track the optimal policy by simply including higher-order interactions. For example, the $K=2\rightarrow 3$ phase transition observed between $\gamma=0.7$ and $\gamma=0.8$ for $|S|=6$ corresponds to the ground state policy switching from the one in Fig.~\ref{roomba_asym}a to the policy in Fig.~\ref{roomba_asym}b.

\begin{figure}
\includegraphics[width=\linewidth]{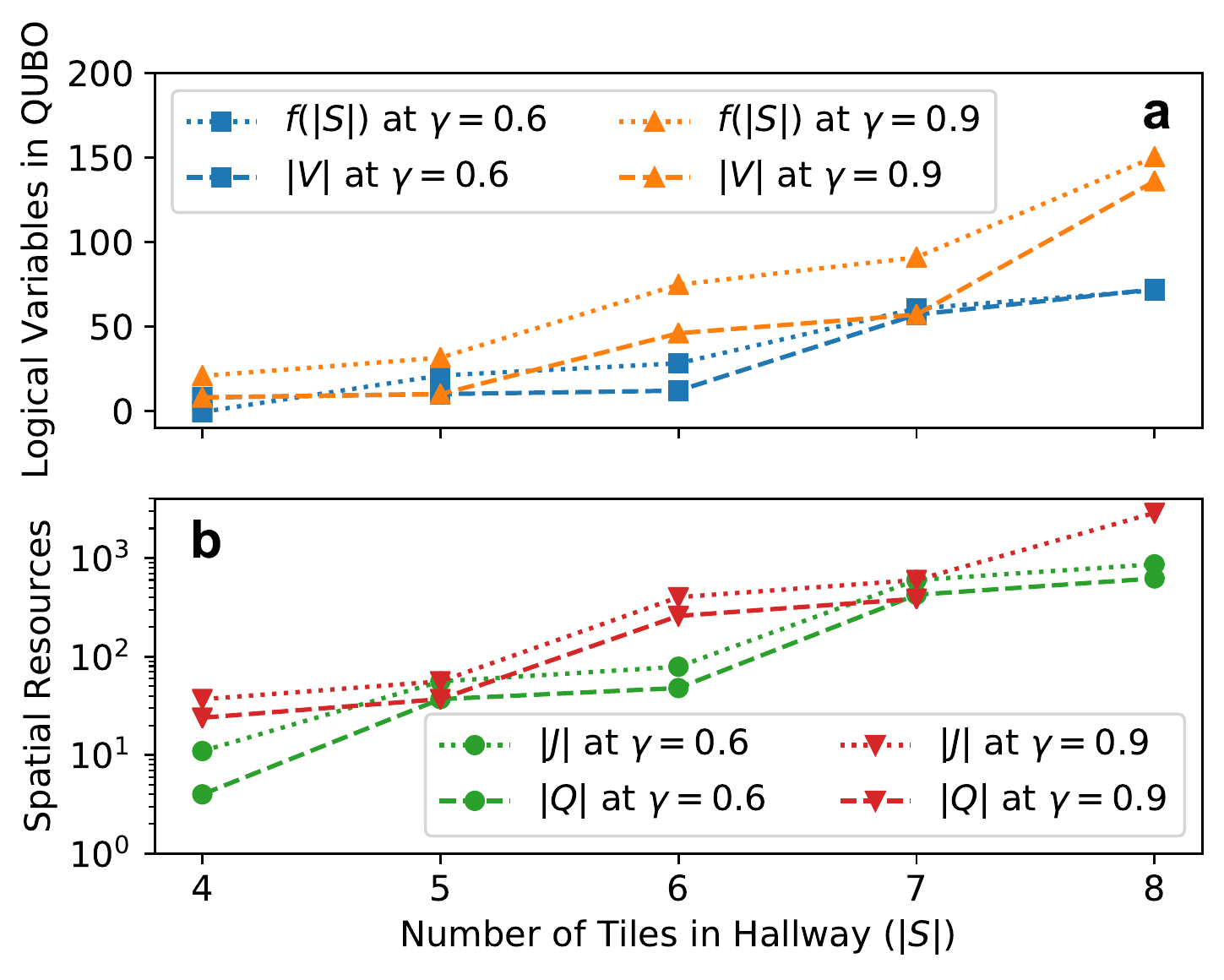}
\caption{Scaling of spatial resources for quantum annealing. \textbf{a} Number of logical qubits ($|V|$, dashed lines for overall trend) needed to encode the truncated Hamiltonian at quadratic order as a function of system size ($|S|$). Dotted lines represent the fitted theoretical scaling function for order reduction by substitution, $f(|S|) = 3\gamma |S \times A|K-25\gamma$. Orange curves with triangles denote $\gamma=0.9$ while blue curves with squares denote $\gamma=0.6$. \textbf{b} Dotted lines represent the number of coefficients ($|J|$) needed to represent the truncated, order-reduced Hamiltonian. Dashed lines represent the number of physical qubits ($|Q|$) required to minor embed the truncated, order-reduced Hamiltonian on the D-Wave 2000Q (16,4)-Chimera hardware graph. Red curves with inverted triangles are at $\gamma=0.9$ while green curves with circles are at $\gamma=0.6$. \label{spatial_scaling}}
\end{figure}

We now turn to an accounting of physical resources required to simulate our Hamiltonian on quantum hardware. In its current form, i.e. before the substitution to qubit operators $\pi_{sa} \rightarrow (\hat{\mathds{1}}_{sa}-\hat{Z}_{sa})/2$, Eq.~\ref{eq:bool_Ham} is in the multilinear polynomial form of a pseudo-Boolean classical cost function. There is a useful result from combinatorial optimization that an arbitrary pseudo-Boolean function with order (degree) $K$ can be reduced to a quadratic-order polynomial in polynomial time \cite{rosenberg1975reduction}. There are a variety of approaches to polynomial order reduction, which typically require the introduction of ancillary variables \cite{dattani2019quadratization}. Here we discuss the most general but worst-scaling technique for order reduction in order to place upper bounds on the quantum hardware required to solve a general MDP. This technique is referred to as substitution.

Substitution for order reduction was first introduced by Rosenberg, and it is predicated on the observation that for three binary variables $x,y,z \in \{0,1\}$, relationships such as
\begin{equation}
xy = (\neq) \: z \: \: \: \mbox{iff} \: \: \: xy - 2xz -2yz+3z = (>) \: 0
\end{equation}
hold \cite{rosenberg1975reduction,boros2002pseudo}. Therefore, one may rewrite a product of two variables $xy$ as a single variable $z$ while at the same time introducing the penalty function
\begin{equation}\label{OR_penalty}
P = M_{OR} (xy - 2xz -2yz+3z)
\end{equation}
into the problem Hamiltonian to enforce $xy=z$, where $M_{OR}$ is the strength of the penalty function. If each of $|S \times A|$ variables occurs in Hamiltonian terms with at most $K-1$ other binary variables, then the number of additional qubits needed to quadratize the problem scales as $O(|S \times A|K)$, where $|S \times A|$ is the magnitude of the MDP state-action space and $K$ is the order at which Eq.~\ref{eq:bool_Ham} is truncated \cite{fix2011graph}. Fig.~\ref{spatial_scaling}a shows the number of logical qubits ($|V|$) required to simulate the truncated, order-reduced Hamiltonian as a function of system size for two different discount factors, $\gamma=0.6$ (blue squares) and $\gamma=0.9$ (orange triangles). Dashed lines connect points calculated by using the D-Wave \textit{make\_quadratic} utility while dotted lines connect points calculated by the function $f(|S|) = 3\gamma |S\times A|K-25\gamma$, which in the asymptotic limit corroborates the $O(|S \times A|K)$ scaling of order reduction by substitution. This scaling is general for a fixed $\gamma$ and independent of the hardware platform used. The dependence of $|V|$ on $\gamma$ is a result of the fact that, on average, for larger discount factors more terms in Eq.~\ref{eq:bool_Ham} need to be taken into account in order to find the correct policy. Once order-reduced, these terms manifest as additional logical variables.

Fig.~\ref{spatial_scaling}b shows the scaling of computational resources required to simulate the truncated, order-reduced Hamiltonian as a function of system size on the D-Wave 2000Q processor in particular for $\gamma=0.6$ (green circles) and $\gamma=0.9$ (red inverted triangles). Dotted lines connect points that denote the number of parameters ($|J|$) of the form in Eq.~\ref{eq:coup_coeffs} needed to specify the Hamiltonian after order reduction while dashed lines connect points that denote the number of physical qubits ($|Q|$) required to simulate the order-reduced Hamiltonian after being minor-embedded on the D-Wave (16,4)-Chimera hardware graph using the \textit{find\_embedding} utility. The \textit{find\_embedding} utility was run ten times with the best resultant embedding taken. Note that the point $|Q|(|S|=8,\gamma=0.9)$ is missing since $|J|(|S|=8,\gamma=0.9)=2,740$ and so the order-reduced Hamiltonian cannot be embedded on the $2,048$ qubit processor.

As observed in previous work, $|J|$ and $|Q|$ track closely to one another over multiple orders of magnitude \cite{jones2020computational}. In addition, at $\gamma=0.6$ and $\gamma=0.9$ as shown, $|Q|$ and $|V|$ both increase by roughly two orders of magnitude between $|S|=4$ and $|S|=8$. This scaling similarity sits well within the worst-case asymptotic $O(|V|^2)$ scaling for embedding a general complete graph ($K_{|V|}$) into a Chimera graph using logical chains \cite{lucas2019hard}. Fig.~\ref{spatial_scaling}b therefore shows that in addition to being amenable to the type of quantum parallelism exhibited by conventional quantum optimization hardware and heuristics, Eq.~\ref{eq:bool_Ham} may suffer from a less severe curse of dimensionality when compared to the classical MDP.

The process of order reduction is necessary in order to solve for the ground state of Eq.~\ref{eq:bool_Ham} on conventional quantum annealers since execution of an annealing schedule is beholden to representing qubit-qubit interactions on the native two-qubit coupling in hardware \cite{boothby2019next}. By contrast, gate-model quantum computers can decompose higher-order interactions via gate compilation into products of single and two-qubit gates \cite{nielsen2002quantum}. Specifically, after truncating and elevating Eq.~\ref{eq:bool_Ham} to an operator and exponentiating it, we find that for a quantum approximate optimization algorithm (QAOA)-like unitary with variational parameter $\beta$, we have
\begin{align}\label{eq:unitary}
e^{-i\beta \hat{H}} &= \prod_{k=0}^{K} \prod_{\mu_1 \ldots \mu_k=1}^{|S\times A|} \Bigg[ \hat{\mathds{1}}_{\mu_1} \otimes \ldots \otimes \hat{\mathds{1}}_{\mu_k} \nonumber \\
&+ \Big(e^{i\beta J_{\mu_1\ldots \mu_k}}-1 \Big) \hat{\pi}_{\mu_1} \otimes \ldots \otimes \hat{\pi}_{\mu_k} \Bigg].
\end{align}
Eq.~\ref{eq:unitary} is equivalent to a product of $(k-1)$-qubit controlled-PHASE $(e^{i\beta J_{\mu_1\ldots \mu_k}})$ gates. Without introducing any additional ancillary qubits, a general N-qubit controlled-unitary may be decomposed with $O(2^N)$ two-qubit gates \cite{barenco1995elementary}. Therefore, without exploiting any hardware-specific implementations, we estimate that for a QAOA of depth $p$, the contribution of Eq.~\ref{eq:unitary} to the gate volume scales at worst as $O(p2^{K|S\times A|^K})$. However, either by introducing ancillary qubits or by hardware specific implementations, scaling of the gate volume can be reduced to $O(pK|S \times A|)$
or better \cite{kumar2013direct}. Keeping in mind that the hardware resources required to solve Eq.~\ref{eq:bool_Ham} are very much architecture dependent, we now turn to a particular experimental assessment of the time complexity of solving the K-spin Hamiltonian by classical simulated annealing and the variant of quantum annealing (transverse field Ising model) offered by the D-Wave 2000Q processor \cite{kadowaki1998quantum}.

\section{Time Complexity of K-spin MDP \label{Time}}

\subsection{Quantum Annealing}

\begin{figure}
\includegraphics[width=\linewidth]{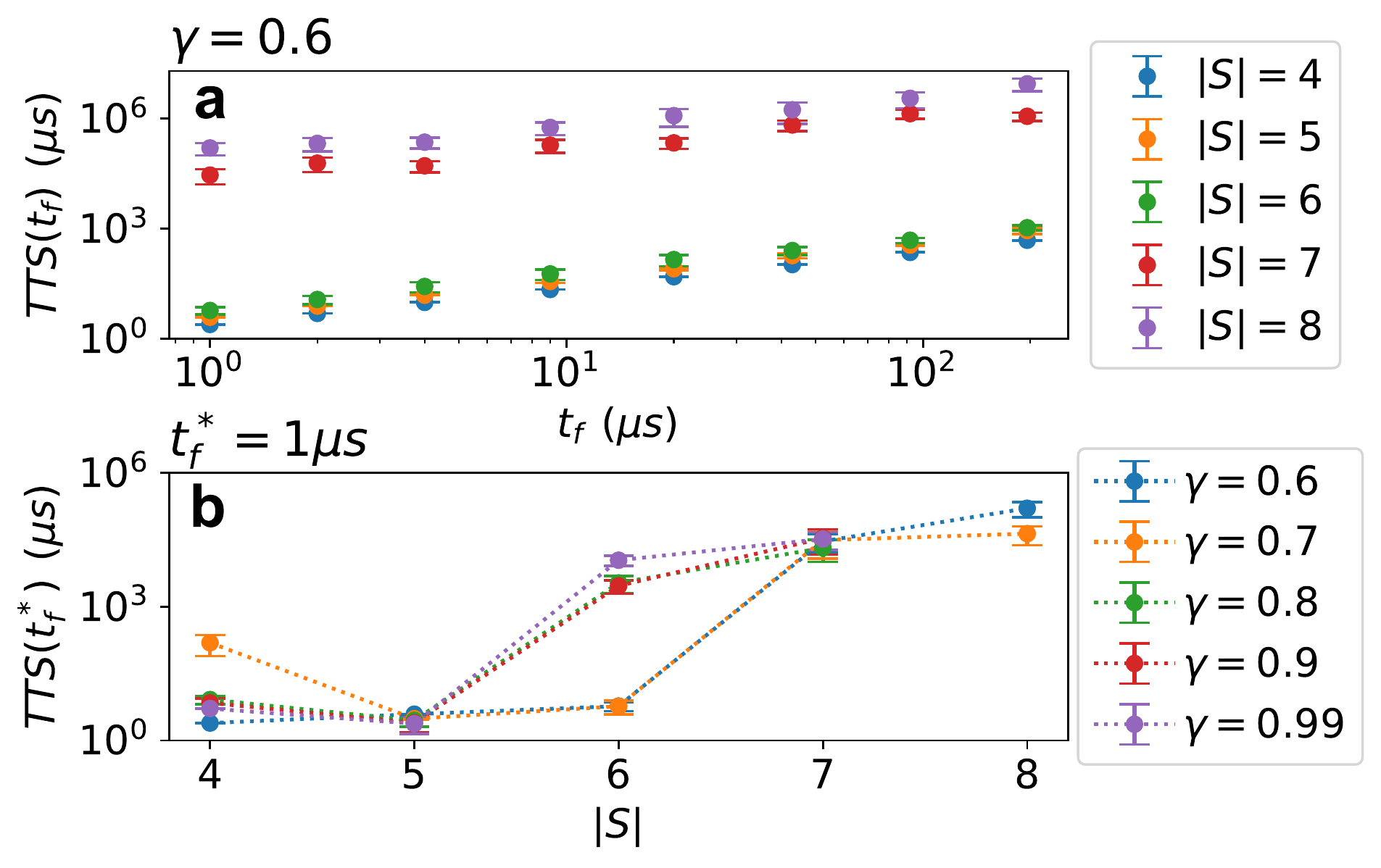}
\caption{Time to solution ($TTS$) for quantum annealing. \textbf{a} $TTS$ as a function of annealing time ($t_f$) for different system sizes ($|S|$) at $\gamma=0.6$. Optimal $TTS$ occurs at (or below) $t_f^*=1 \mu s$ for all system sizes, which is representative of all other values for gamma. Error bars represent one standard deviation from the mean. \textbf{b} Optimal time-to-solution ($TTS(t_f^*)$) as a function of system size ($|S|$) for different values of $\gamma$. Error bars represent one standard deviation from the mean. \label{temporal_scaling}}
\end{figure}

A common and well-defined time-to-solution ($TTS$) metric for the D-Wave 2000Q quantum annealer is
\begin{equation}\label{eq:QA_TTS}
TTS(t_f) = t_f \frac{\ln(1-p_d)}{\ln(1-p_s(t_f))},
\end{equation}
where here $t_f \in [1\mu s,200\mu s]$ is the time chosen over which each annealing schedule is carried out, $p_d$ is a desired success probability often taken to be $0.99$, and for a given annealing time, $p_s(t_f)$ is the probability of success for a single run of the annealer \cite{denchev2016computational,albash2018demonstration}. For both quantum annealing calculations and simulated annealing calculations, we set $M=3$ and $M_{OR}=5$. These values are chosen such that the penalty functions for conditional probability conservation of the policy (second line in Eq.~\ref{eq:bool_Ham}) and consistency terms from order-reduction (Eq.~\ref{OR_penalty}), respectively, are costly to violate. We note that the particular values of $M=3$ and $M_{OR}=5$ are somewhat arbitrary in that any $M \approx M_{OR} \gtrsim 2$ will make the attendant penalty functions sufficiently costly to violate. In each problem instance, the chain strength to enforce the logical consistency of embedded variables was set to $J_{chain} = 1.5 |J|_M$, where $|J|_M$ is the largest magnitude coupling term in the order-reduced QUBO. This value was chosen so that the fraction of broken logical chains was less than 0.001. 

Fig.~\ref{temporal_scaling}a shows a representative set of curves for $TTS(t_f)$ as a function of annealing time at $\gamma=0.6$. At each system size, it is clear that the optimal time-to-solution $TTS(t_f^*)$ occurs at (or below) $t_f^*=1\mu s$, which is the shortest annealing time available on the processor used. This is also the case for all other values of $\gamma$. As such, Fig.~\ref{temporal_scaling}b shows $TTS(t_f^*)$ as a function of $|S|$ for all values of $\gamma$. While $TTS(t_f^*)$ actually appears to decrease between $|S|=4$ and $5$ for $\gamma \geq 0.7$, likely due to the boundary condition effect for small $|S|$ mentioned in Sec.~\ref{Hardware}, $TTS(t_f^*)$ between the rest of the values of $|S|$ appears to follow a sort piecewise polynomial scaling where large jumps in $TTS(t_f^*)$ are concurrent with large jumps in the number of physical qubits ($|Q|$) in Fig.~\ref{spatial_scaling}b. Indeed, the dependence of $TTS(t_f^*)$ on $\gamma$ is a direct result of the dependence of $|V|$ and therefore $|Q|$ on $\gamma$. However, that $TTS(t_f^*)$ ranges over five orders of magnitude while stepping through just a twofold increase in $|S|$ speaks to asymptotic exponential scaling of the solution of Eq.~\ref{eq:bool_Ham} by quantum annealing on the D-Wave 2000Q, in clear analogy with solving $k$-regular $k$-XORSAT-- another polynomial-time solvable problem-- by quantum annealing \cite{farhi2012performance,patil2019obstacles,bapst2013quantum}. 

\subsection{Simulated Annealing}

\begin{figure}
\includegraphics[width=\linewidth]{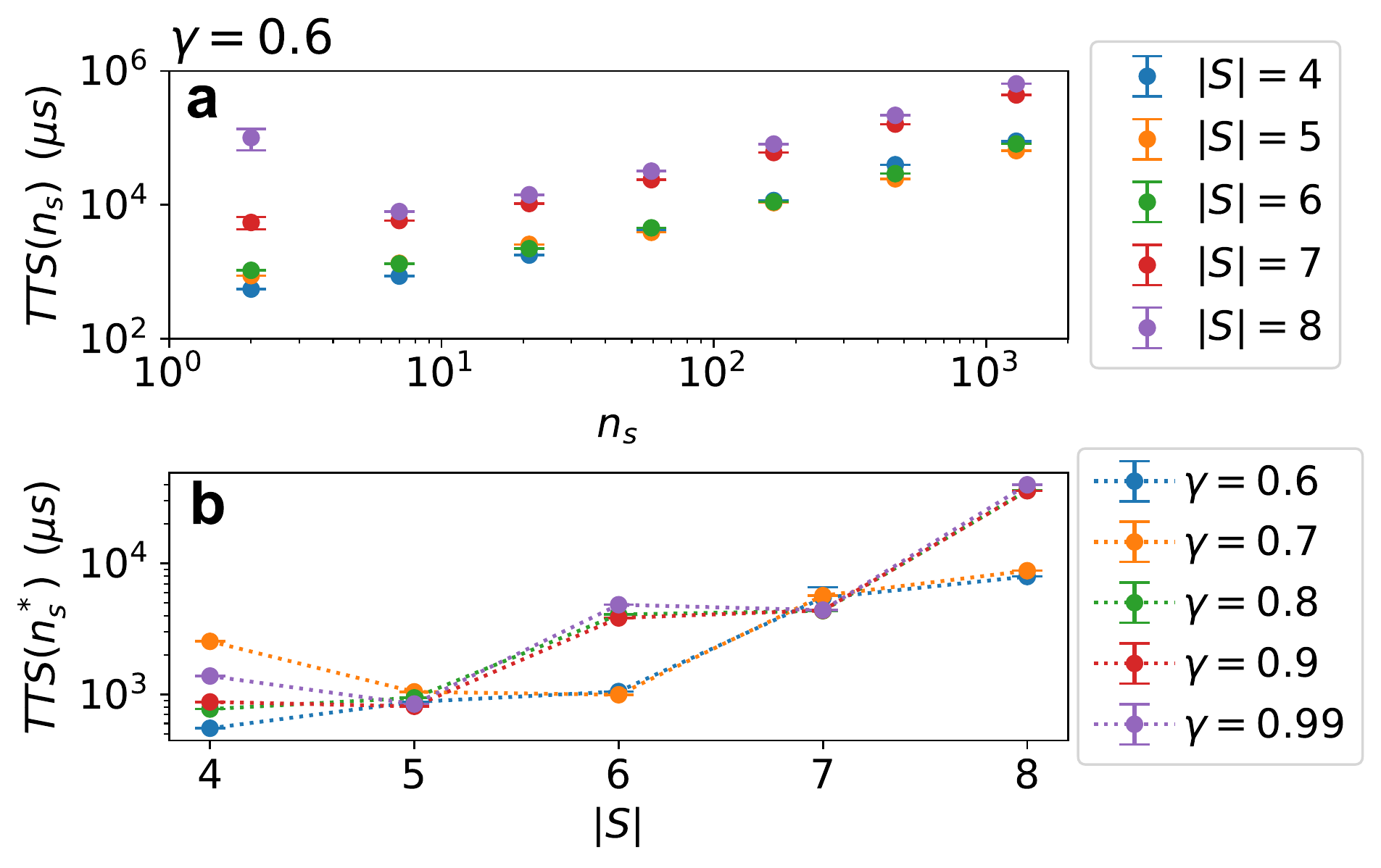}
\caption{Time to solution ($TTS$) for simulated annealing. \textbf{a} $TTS$ as a function of number of Monte Carlo sweeps ($n_s$) for different system sizes ($|S|$) at $\gamma=0.6$. Optimal $TTS$ now generally depends upon both $|S|$ and $\gamma$. Error bars represent one standard deviation from the mean. \textbf{b} Optimal time-to-solution ($TTS(n_s^*)$) as a function of system size ($|S|$) for different values of $\gamma$. Error bars represent one standard deviation from the mean with most being smaller than the marker size. \label{SA_temporal_scaling}}
\end{figure}

An analog to Eq.~\ref{eq:QA_TTS} for simulated annealing is 
\begin{equation}
TTS(t_a) =  t_a \frac{\ln(1-p_d)}{\ln(1-p_s(t_a))},
\end{equation}
where for large $n_s \cdot N$, $t_a = n_s \cdot N / f$, but in general $t_a = t_a(n_s)$ \cite{isakov2015optimised}. Here, $n_s$ is the total number of Monte Carlo sweeps (that is, the number of times all spins see an attempted update at a given temperature times the number of temperature steps) involved in a particular annealing run, $N$ is the number of logical spins needed to represent the system under consideration, and $f$ is the attempted spin update frequency. Note that since $f$ is a fixed property of the implementation and classical hardware being used across all problem instances and since $N=|V|$ is fixed by the problem instance, we consider $TTS(n_s)$ in Fig.~\ref{SA_temporal_scaling}a. In this plot, for $\gamma=0.6$ and $|S|=8$, $n^*_s \approx 7$ while for all other system sizes at this discount factor $n_s \approx 2$. In general for the simulated annealing algorithm presented here $n_s^*$ will depend upon both $|S|$ and $\gamma$. The simulated annealing implementation used here was the D-Wave \textit{SimulatedAnnealingSampler} with default, linearly interpolated $\beta=1/k_B T$ values run on a single 2.5 GHz Intel Core i5 processor with 16 GB of 1600 MHz DDR3 memory (source code at \cite{SA_source}). Taken together with the results in Fig.~\ref{temporal_scaling}b, the results in Fig.~\ref{SA_temporal_scaling}b support the likely asymptotically exponential time complexity for solving the MDP as a spin Hamiltonian.

\section{Conclusion \label{Conclusion}}
Here we have established a direct link between the Markov decision process formalism of reinforcement learning and a quantum Hamiltonian amenable to solution on current-generation and near-term quantum hardware. Our derivation demonstrates that some problems that resist parallelization classically may become manifestly parallel when recast into the quantum model of computation. Despite this parallelization, the K-spin Hamiltonian likely scales exponentially in time when solved using conventional simulated and quantum annealing resources. Moreover, spatial fully-connected quantum computational resources scale polynomially in the size of the state space, while limited hardware connectivity induces a polynomial growth in spatial resources that is at most quadratically worse than the case for fully-connected resources. Despite these limitations, our K-spin Hamiltonian provides another previously unexplored inroad to quantum formulations of full reinforcement learning as well as more difficult variants of the MDP such as POMDP.

\begin{acknowledgments}
This work was authored in part by the National Renewable Energy Laboratory (NREL), operated by Alliance for Sustainable Energy, LLC, for the U.S. Department of Energy (DOE) under Contract No. DE-AC36-08GO28308. This work was supported by the Laboratory Directed Research and Development (LDRD) Program at NREL. The views expressed in the article do not necessarily represent the views of the DOE or the U.S. Government. The U.S. Government retains and the publisher, by accepting the article for publication, acknowledges that the U.S. Government retains a nonexclusive, paid-up, irrevocable, worldwide license to publish or reproduce the published form of this work, or allow others to do so, for U.S. Government purposes. This research used Ising, Los Alamos National Laboratory's D-Wave quantum annealer.  Ising is supported by NNSA's Advanced Simulation and Computing program. The authors would like to thank Scott Pakin and Denny Dahl. This material is based in-part upon work supported by the National Science Foundation under Grant No. PHY-1653820.
\end{acknowledgments}


%

\end{document}